\documentclass{article}

\usepackage{cite}
\usepackage{graphicx}
\usepackage{dcolumn}

\begin{document}

\date{}
\title{On the symmetry of four particles in a one-dimensional box with harmonic
interaction}
\author{Francisco M. Fern\'{a}ndez \thanks{%
E-mail: fernande@quimica.unlp.edu.ar} \\
INIFTA (CONICET, UNLP), Divisi\'on Qu\'imica Te\'orica\\
Blvd. 113 S/N, Sucursal 4, Casilla de Correo 16, 1900 La Plata, Argentina}
\maketitle

\begin{abstract}
We show that a system of four particles in a one-dimensional box with a
two-particle harmonic interaction can by described by means of the symmetry
point group $O_h$. Group theory proves useful for the discussion of both the
small-box and large-box regimes. We apply perturbation theory and obtain the
corrections of first order for the lowest states. We carry out a simple
Rayleigh-Ritz variational calculation with basis sets adapted to the
symmetries of the system. We also obtain alternative variational results for
the first three lowest energy levels that are more suitable for larger box
sizes.
\end{abstract}

\section{Introduction}

\label{sec:intro}

During the last decades there has been great interest in the model of a
harmonic oscillator confined to boxes of different shapes, sizes and
dimensions. Such model has been suitable for the study of several physical
problems ranging from dynamical friction in star clusters to magnetic
properties of solids and impurities in quantum dots (see\cite{AF10,AF15} for
a review of the relevant literature on the subject). In addition to it,
systems of few identical particles in one dimension have proved to exhibit a
rich phenomenological structure resembling that of realistic systems\cite
{H12,H14,H16a,H16b}. In two recent papers Amore and Fern\'{a}ndez discussed
the problems posed by two\cite{AF10} and three\cite{AF15} particles confined
in a one-dimensional box with impenetrable walls that interact through
harmonic forces. They found that a straightforward application of group
theory considerably facilitates the analysis of the solutions to the
Schr\"{o}dinger equation. In particular, an accurate Rayleigh-Ritz
variational calculation revealed that the energies of the three-particle
model as functions of the box length exhibit a most interesting pattern of
avoided crossings between pairs of states of the same symmetry\cite{AF15}.

The purpose of this paper is to discuss the case of four particles in a one
dimensional box that also interact through harmonic forces. The reason for
choosing such interaction is that it has proved quite useful in the past
(see the references in\cite{AF15}) and because the calculation of the matrix
elements of the resulting potential is quite simple. In section~\ref
{sec:model} we introduce the model, section~\ref{sec:small-box} shows a
perturbation approach to the small-box regime and a Rayleigh-Ritz
calculation with symmetry-adapted basis sets, section~\ref{sec:large-box}
describes the large-box limit, in section~\ref{sec:variational} we discuss a
simple variational calculation that is more suitable for larger box sizes
and in section~\ref{sec:conclusions} we draw conclusions. There is also an
Appendix outlining the construction of the projection operators used in all
the calculations just described.

\section{Four particles in a one-dimensional box}

\label{sec:model}

We first consider $N$ interacting point particles of mass $m$ in a
one-dimensional box of length $L=2a$ with Hamiltonian
\begin{equation}
H=-\frac{\hbar ^{2}}{2m}\sum_{i=1}^{N}\frac{\partial ^{2}}{\partial x_{i}^{2}%
}+\sum_{i=1}^{N-1}\sum_{j=i+1}^{N}W(|x_{i}-x_{j}|),  \label{eq:H_ND}
\end{equation}
where $x_{i}$ is the coorinate of the $i$-th particle. The boundary
conditions are determined by the impenetrable walls of the box
\begin{equation}
\Psi (x_{1},x_{2},\ldots ,x_{i}=\pm a,\ldots ,x_{N})=0,\;i=1,2,\ldots ,N.
\label{eq:BC_ND}
\end{equation}
The Hamiltonian operator is invariant under the $N!$ permutations of the
particle coordinates as well as under parity inversion $\mathbf{x}%
\rightarrow -\mathbf{x}$. The $2N!$ $N\times N$ matrices that produce all
the permutations of the sets $\{x_{1},x_{2},\ldots ,x_{N}\}$ and $%
\{-x_{1},-x_{2},\ldots ,-x_{N}\}$ form a group given by the product $%
S_{N}\otimes O(1)$\cite{H62,T64}. When $N=2$ the group is also called $%
C_{2h} $ (in principle we can also use $D_{2}$ or $C_{2v}$ that are
isomorphic to $C_{2h}$) and when $N=3$ we can resort to either $D_{3d}$ or $%
D_{3h}$\cite{C90}. Both cases have already been treated by group theory in
earlier papers\cite{AF10,AF15}.

In order to solve the Schr\"{o}dinger equation it is convenient to define
the dimensionless particle coordinates $q_{i}=x_{i}/a$ and the dimensionless
Hamiltonian
\begin{equation}
H_{d}=\frac{2ma^{2}}{\hbar ^{2}}H=-\sum_{i=1}^{N}\frac{\partial ^{2}}{%
\partial q_{i}^{2}}+\lambda \sum_{i=1}^{N-1}\sum_{j=i+1}^{N}w(|q_{i}-q_{j}|),
\label{eq:H_d(ND)}
\end{equation}
where $\lambda w(|q_{i}-q_{j}|)=2ma^{2}W(a|q_{i}-q_{j}|)/\hbar ^{2}$. The
boundary conditions for the eigenfunctions $\psi $ of this operator now
become
\begin{equation}
\psi (q_{1},q_{2},\ldots ,q_{i}=\pm 1,\ldots ,q_{N})=0,\;i=1,2,\ldots ,N.
\label{eq:BC_ND_dimensionless}
\end{equation}
From now on we only consider the dimensionless Hamiltonian (\ref{eq:H_d(ND)}%
) and omit the subscript $d$. In order to facilitate the numerical
calculations we choose a harmonic interaction of the form $W\left( \left|
x_{i}-x_{j}\right| \right) =\frac{k}{2}\left( x_{i}-x_{j}\right) ^{2}$ that
leads to $\lambda =$ $ma^{2}k/\hbar ^{2}$ and $%
w(|q_{i}-q_{j}|)=(q_{i}-q_{j})^{2}$.

For $N=4$ our problem just reduces to solving the Schr\"{o}dinger equation
for the operator
\begin{eqnarray}
H=- &&\left( \frac{\partial ^{2}}{\partial q_{1}^{2}}+\frac{\partial ^{2}}{%
\partial q_{2}^{2}}+\frac{\partial ^{2}}{\partial q_{3}^{2}}+\frac{\partial
^{2}}{\partial q_{4}^{2}}\right) +  \nonumber \\
&+&\lambda \left[ \left( q_{1}-q_{2}\right) ^{2}+\left( q_{1}-q_{3}\right)
^{2}+\left( q_{1}-q_{4}\right) ^{2}+\left( q_{2}-q_{3}\right) ^{2}\right.
\nonumber \\
&&\left. +\left( q_{2}-q_{4}\right) ^{2}+\left( q_{3}-q_{4}\right)
^{2}\right] ,  \label{eq:H_dimensionless}
\end{eqnarray}
with the boundary conditions
\begin{equation}
\psi (\pm 1,q_{2},q_{3},q_{4})=\psi (q_{1},\pm 1,q_{3},q_{4})=\psi
(q_{1},q_{2},\pm 1,q_{4})=\psi (q_{1},q_{2},q_{3},\pm 1)=0.
\label{eq:box_bc}
\end{equation}
In this case the group $S_{4}\otimes O(1)$ is isomorphic to $O_{h}$ and in
this paper we resort to the character table of the latter group\cite{C90}.

\section{Small-box-regime}

\label{sec:small-box}

When $\lambda $ is sufficiently small (sufficiently small box size $L$) we
can estimate the energy levels by means of perturbation theory. The
Schr\"{o}dinger equation is exactly solvable when $\lambda =0$ and the
eigenvalues and eigenfunctions of $H_{0}=H(\lambda =0)$ are
\begin{eqnarray}
E_{n_{1}n_{2}n_{3}n_{4}}^{(0)} &=&\frac{\pi ^{2}}{4}\left(
n_{1}^{2}+n_{2}^{2}+n_{3}^{2}+n_{4}^{2}\right)
,\;n_{1},n_{2},n_{3},n_{4}=1,2,\ldots  \nonumber \\
\psi _{n_{1}n_{2}n_{3}n_{4}}^{(0)}(q_{1},q_{2},q_{3},q_{4}) &=&\phi
_{n_{1}}(q_{1})\phi _{n_{2}}(q_{2})\phi _{n_{3}}(q_{3})\phi _{n_{4}}(q_{4})
\nonumber \\
\phi _{n}(q) &=&\sin \frac{n\pi (q+1)}{2}.  \label{eq:H0_eigenval_eigenfun}
\end{eqnarray}
Note that $\phi _{n}(-q)=(-1)^{n-1}\phi _{n}(q)$.

In order to facilitate the discussion of the results we introduce the
notation $\{a,b,c,d\}_{P}$ to indicate the set of all distinct permutations
of four elements that may be either coordinates or quantum numbers. For
example, each 4-tuple of quantum numbers in the set $%
\{n_{1},n_{2},n_{3},n_{4}\}_{P}$ leads to the same unperturbed energy $%
E_{n_{1}n_{2}n_{3}n_{4}}^{(0)}$. We may eventually add accidental \textit{%
Pythagorean} degeneracies of the form $%
m_{1}^{2}+m_{2}^{2}+m_{3}^{2}+m_{4}^{2}=n_{1}^{2}+n_{2}^{2}+n_{3}^{2}+n_{4}^{2}
$, where $\left( m_{1},m_{2},m_{3},m_{4}\right) \notin
\{n_{1},n_{2},n_{3},n_{4}\}_{P}$\cite{F13b}. If such energy level is $g$%
-fold degenerate then the perturbation corrections of first order to the
eigenfunctions will be of the form
\begin{equation}
\psi ^{(1)}=\sum_{j=1}^{g}c_{j}^{(1)}\psi _{j}^{(0)},  \label{eq:psi^(1)}
\end{equation}
where $j$ denotes a 4-tuple $(n_{1},n_{2},n_{3},n_{4})$. The coefficients $%
c_{j}^{(1)}$ are solutions to the secular equation
\begin{equation}
\left( \mathbf{H}^{\prime }-E^{(1)}\mathbf{I}\right) \mathbf{c}^{(1)}=0,
\label{eq:secular}
\end{equation}
where $\mathbf{H}^{\prime }$ is the $g\times g$ matrix of the perturbation $%
H^{\prime }=H-H_{0}$ in the set of degenerate eigenfunctions $\left\{ \psi
_{j}^{(0)},j=1,2,\ldots ,g\right\} $, $\mathbf{I}$ is the $g\times g$
identity matrix, $\mathbf{c}^{(1)}$ is a column vector with elements $%
c_{j}^{(1)}$ and $E^{(1)}$ is one of the $g$ roots of the secular
determinant $\left| \mathbf{H}^{\prime }-E^{(1)}\mathbf{I}\right| =0$.

In order to determine the symmetry of the solution $\psi ^{(1)}$, which
provides a suitable label for the corresponding root $E^{(1)}$, we apply the
projection operators $P_{S}$ associated to the irreducible representations
(irreps) $S$ of the group $O_{h}$\cite{C90}. The result is well known to be $%
P_{S}\psi ^{(1)}=\psi ^{(1)}$ if $\psi ^{(1)}$ is a basis for the irrep $S$
or $P_{S}\psi ^{(1)}=0$ otherwise. The construction of the projection
operators is outlined in the Appendix. In this way we obtain the following
results for the first energy levels:

\begin{equation}
E_{1A_{1g}}=\pi ^{2}+\frac{4\left( \pi ^{2}-6\right) }{\pi ^{2}}\lambda
+\ldots ,  \label{eq:E1g}
\end{equation}
\begin{eqnarray}
E_{1A_{1u}} &=&\frac{7\pi ^{2}}{4}+\frac{216\pi ^{4}-1053\pi ^{2}-4096}{%
54\pi ^{4}}\lambda +\ldots ,  \nonumber \\
E_{1T_{2u}} &=&\frac{7\pi ^{2}}{4}+\frac{648\pi ^{4}-3159\pi ^{2}+4096}{%
162\pi ^{4}}\lambda +\ldots ,  \label{eq:E1u}
\end{eqnarray}

\begin{eqnarray}
E_{2A_{1g}} &=&\frac{5\pi ^{2}}{2}+\frac{324\pi ^{4}-1215\pi ^{2}-8192}{%
81\pi ^{4}}\lambda +\ldots ,  \nonumber \\
E_{1T_{2g}} &=&\frac{5\pi ^{2}}{2}+\frac{4\pi ^{2}-15}{\pi ^{2}}\lambda
+\ldots ,  \nonumber \\
E_{1E_{g}} &=&\frac{5\pi ^{2}}{2}+\frac{324\pi ^{4}-1215\pi ^{2}+4096}{81\pi
^{4}}\lambda +\ldots ,  \label{eq:E2g}
\end{eqnarray}

\begin{eqnarray}
E_{3A_{1g}} &=&3\pi ^{2}+\frac{4\left( 3\pi ^{2}-14\right) }{3\pi ^{2}}%
\lambda +\ldots ,  \nonumber \\
E_{2T_{2g}} &=&3\pi ^{2}+\frac{4\left( 3\pi ^{2}-14\right) }{3\pi ^{2}}%
\lambda +\ldots ,  \label{eq:E3g}
\end{eqnarray}

\begin{eqnarray}
E_{2A_{1u}} &=&\frac{13\pi ^{2}}{4}+\frac{216\pi ^{4}-567\pi ^{2}-4096}{%
54\pi ^{4}}\lambda +\ldots ,  \nonumber \\
E_{2T_{2u}} &=&\frac{13\pi ^{2}}{4}+\frac{648\pi ^{4}-1701\pi ^{2}+4096}{%
162\pi ^{4}}\lambda +\ldots ,  \label{eq:E2u}
\end{eqnarray}

\begin{eqnarray}
E_{3A_{1u}} &=&\frac{15\pi ^{2}}{4}+\frac{405000\pi ^{4}-1434375\pi
^{2}-8105984}{101250\pi ^{4}}\lambda +\ldots ,  \nonumber \\
E_{3T_{2u}} &=&\frac{15\pi ^{2}}{4}+\frac{405000\pi ^{4}-1434375\pi ^{2}-6144%
\sqrt{424321}-1280000}{101250\pi ^{4}}\lambda +\ldots ,  \nonumber \\
E_{1E_{u}} &=&\frac{15\pi ^{2}}{4}+\frac{405000\pi ^{4}-1434375\pi
^{2}-425984}{101250\pi ^{4}}\lambda +\ldots ,  \nonumber \\
E_{4T_{2u}} &=&\frac{15\pi ^{2}}{4}+\frac{405000\pi ^{4}-1434375\pi ^{2}+6144%
\sqrt{424321}-1280000}{101250\pi ^{4}}\lambda +\ldots ,  \nonumber \\
E_{1T_{1u}} &=&\frac{15\pi ^{2}}{4}+\frac{405000\pi ^{4}-1434375\pi
^{2}+5545984}{101250\pi ^{4}}\lambda +\ldots ,  \label{eq:E3u}
\end{eqnarray}

\begin{equation}
E_{4A_{1g}}=4\pi ^{2}+\frac{2\left( 2\pi ^{2}-3\right) }{\pi ^{2}}\lambda
+\ldots ,  \label{eq:E4g}
\end{equation}

\begin{eqnarray}
E_{5A_{1g}} &=&\frac{9\pi ^{2}}{2}+\frac{202500\pi ^{4}-489375\pi
^{2}-5545984}{50625\pi ^{4}}\lambda +\ldots ,  \nonumber \\
E_{3T_{2g}} &=&\frac{9\pi ^{2}}{2}+\frac{202500\pi ^{4}-489375\pi ^{2}-6144%
\sqrt{53329}-1386496}{50625\pi ^{4}}\lambda +\ldots ,  \nonumber \\
E_{4T_{2g}} &=&\frac{9\pi ^{2}}{2}+\frac{202500\pi ^{4}-489375\pi ^{2}+6144%
\sqrt{53329}-1386496}{50625\pi ^{4}}\lambda +\ldots ,  \nonumber \\
E_{1T_{1g}} &=&\frac{9\pi ^{2}}{2}+\frac{202500\pi ^{4}-489375\pi
^{2}+2772992}{50625\pi ^{4}}\lambda +\ldots ,  \nonumber \\
E_{2E_{g}} &=&\frac{9\pi ^{2}}{2}+\frac{202500\pi ^{4}-489375\pi ^{2}+2772992%
}{50625\pi ^{4}}\lambda +\ldots ,  \label{eq:E5g}
\end{eqnarray}
where $A_{1g}$, $A_{2g}$, $A_{1u}$, and $A_{2u}$ are nondegenerate, $E_{g}$
and $E_{u}$ are two-fold degenerate, and $T_{1g}$, $T_{2g}$, $T_{1u}$ and $%
T_{2u}$ are three-fold degenerate.

We appreciate that the degeneracy of the unperturbed states is partially
removed by the perturbation. The remaining degeneracies are expected to be
broken at higher perturbation orders. One does not expect that such
remaining degeneracies are due to an unknown hidden symmetry\cite{F13b}
because they are rather inconsistent. For example, in the case $%
\{1,1,1,3\}_{P}$ we have $E_{3A_{1g}}=E_{2T_{2g}}$; however, $%
\{1,2,2,3\}_{P} $ leads to $%
E_{5A_{1g}}<E_{3T_{2g}}<E_{4T_{2g}}<E_{2E_{g}}=E_{1T_{1g}}$ where the state $%
A_{1g}$ and the three states $T_{2g}$ are not degenerate as in the preceding
case.

\begin{figure}[]
\begin{center}
\includegraphics[width=9cm]{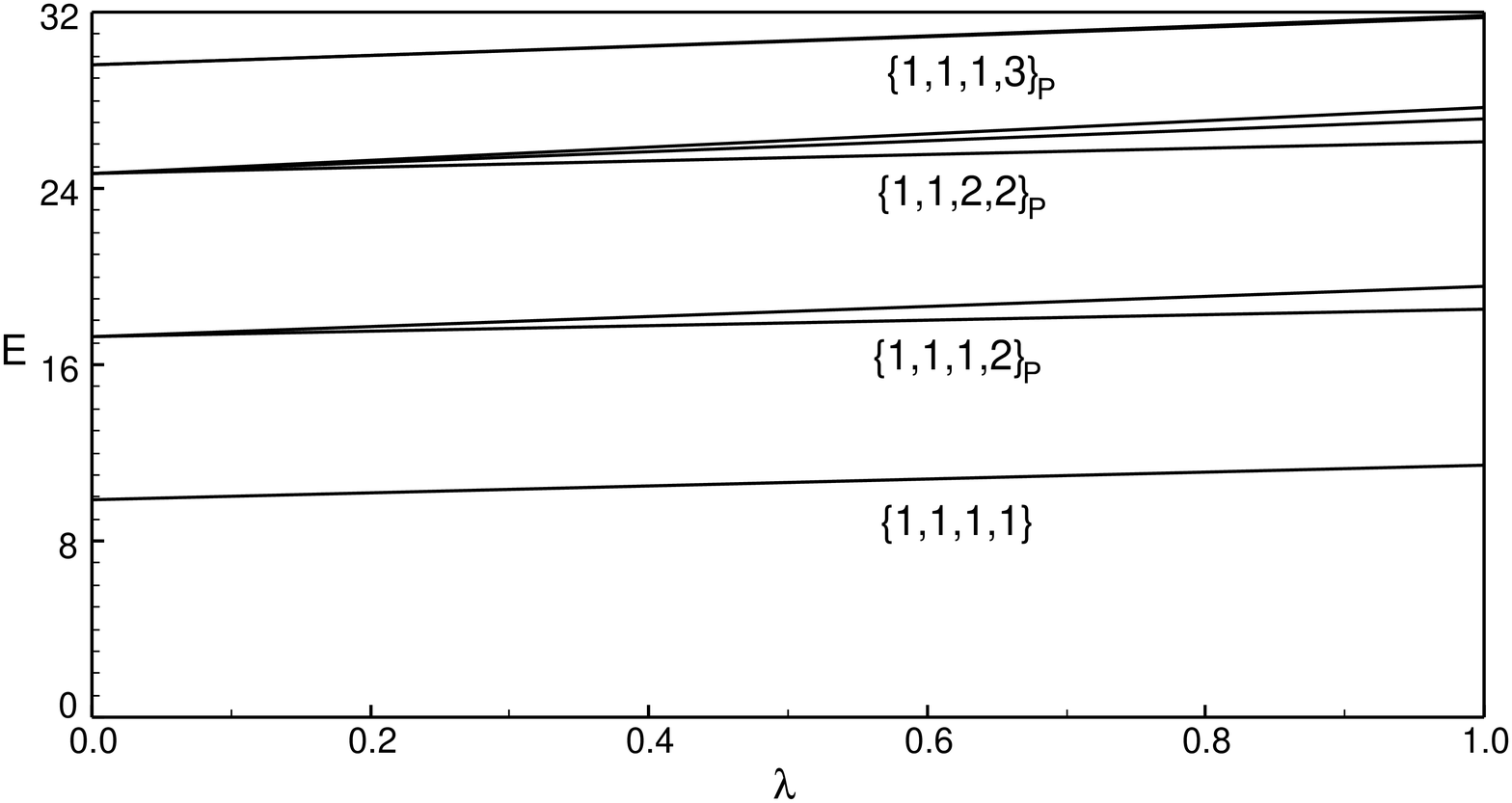} \includegraphics[width=9cm]{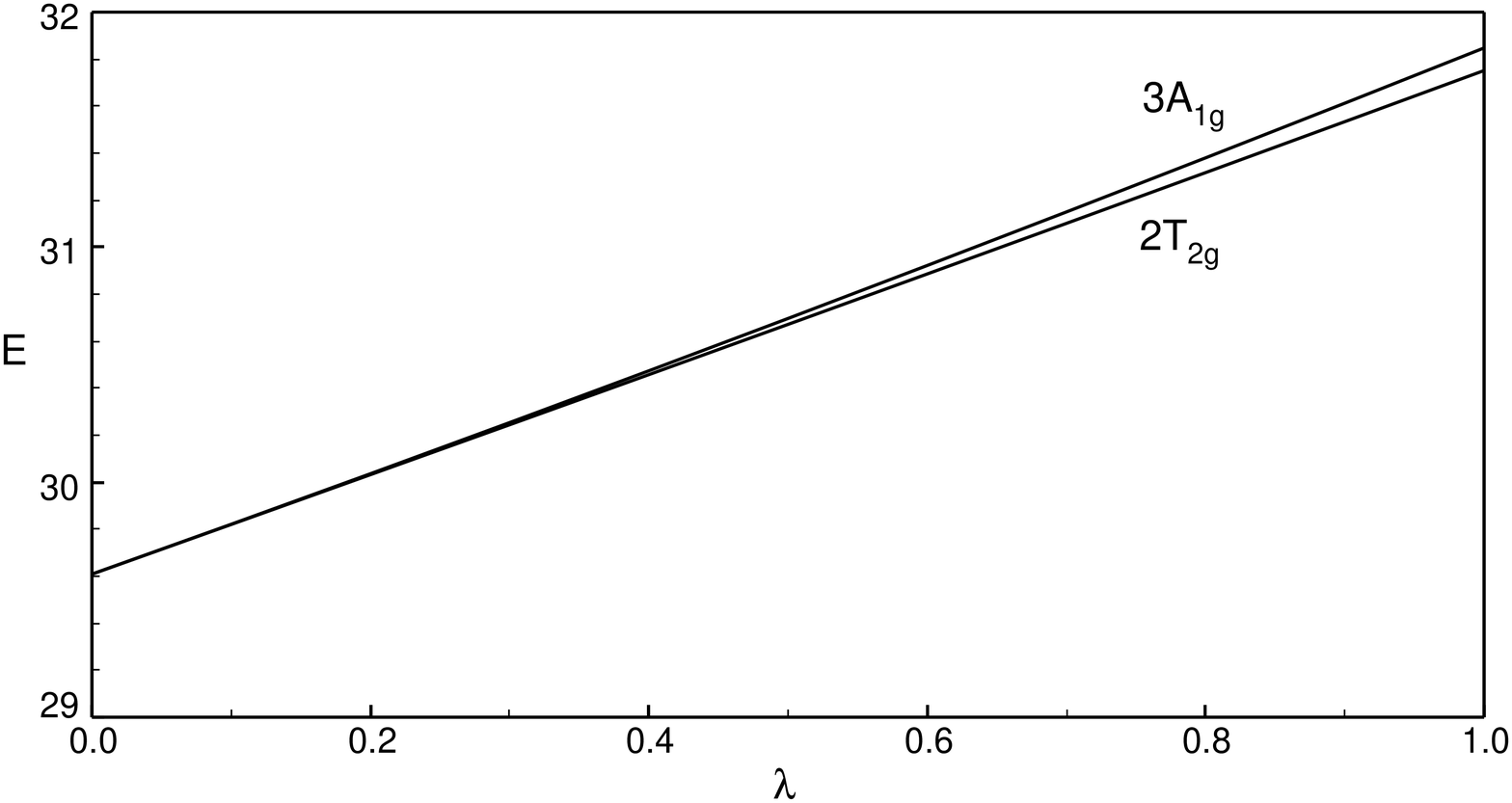}
\end{center}
\caption{Lowest eigenvalues calculated by means of perturbation theory and
the Raleigh-Ritz variational method}
\label{fig:EPTRR}
\end{figure}

Figure~\ref{fig:EPTRR} shows the lowest energy levels in the interval $0\leq
\lambda \leq 1$ where perturbation theory is expected to yield sufficiently
accurate results. We also carried out a simple Rayleigh-Ritz variational
calculation with symmetry-adapted basis sets for $A_{1g}$ and $T_{2g}$
choosing only those functions coming from $\{1,1,1,1\}$, $\{1,1,2,2\}_{P}$
and $\{1,1,1,3\}_{P}$ in order to show the splitting of the levels $%
E_{3A_{1g}}$ and $E_{2T_{2g}}$. These levels appear to be degenerate in the
upper subfigure but the finer scale of the lower one clearly reveals the
splitting that takes place at the second order of perturbation theory ($%
E_{3A_{1g}}>E_{2T_{2g}}$). The symmetry-adapted basis set was constructed by
straightforward application of the projection operators $P_{A_{1g}}$ and $%
P_{T_{2g}}$ to the zeroth-order eigenfunctions (\ref{eq:H0_eigenval_eigenfun}%
) (see \cite{AF10,AF15} for more details).

\section{Large-box-regime}

\label{sec:large-box}

When $L\rightarrow \infty $ ($\lambda \rightarrow \infty $) we have a system
of four unconfined particles with harmonic-pair interaction. In order to
discuss this case it is convenient to define the new coordinates

\begin{eqnarray}
\xi _{1} &=&\frac{\sqrt{2}q_{2}}{2}-\frac{\sqrt{2}q_{1}}{2}  \nonumber \\
\xi _{2} &=&-\frac{\sqrt{6}q_{1}}{6}-\frac{\sqrt{6}q_{2}}{6}+\frac{\sqrt{6}%
q_{3}}{3}  \nonumber \\
\xi _{3} &=&-\frac{\sqrt{3}q_{1}}{6}-\frac{\sqrt{3}q_{2}}{6}-\frac{\sqrt{3}%
q_{3}}{6}+\frac{\sqrt{3}q_{4}}{2}  \nonumber \\
\xi _{4} &=&\frac{q_{1}+q_{2}+q_{3}+q_{4}}{2}  \label{eq:xi(q)}
\end{eqnarray}
because the Hamiltonian operator becomes
\begin{equation}
H=-\left( \frac{\partial ^{2}}{\partial \xi _{1}^{2}}+\frac{\partial ^{2}}{%
\partial \xi _{2}^{2}}+\frac{\partial ^{2}}{\partial \xi _{3}^{2}}+\frac{%
\partial ^{2}}{\partial \xi _{4}^{2}}\right) +4\lambda \left( \xi
_{1}^{2}+\xi _{2}^{2}+\xi _{3}^{2}\right) .  \label{eq:H(xi_j)}
\end{equation}
We appreciate that the center of mass described by the coordinate $\xi _{4}$
moves freely while there is harmonic oscillation along the 3 remaining
coordinates. The eigenvalues and eigenfunctions are expected to be
\begin{eqnarray}
E_{K,n_{1},n_{2},n_{3}} &=&K^{2}+2\sqrt{\lambda }\left(
2n_{1}+2n_{2}+2n_{3}+3\right) ,  \nonumber \\
\psi _{K,n_{1},n_{2},n_{3}}(\xi _{1},\xi _{2},\xi _{3},\xi _{4}) &=&\exp
\left( iK\xi _{4}\right) \chi _{n_{1}}(\xi _{1})\chi _{n_{2}}(\xi _{2})\chi
_{n_{3}}(\xi _{3}),  \nonumber \\
n_{1},n_{2},n_{3} &=&0,1,\ldots ,\;-\infty <K<\infty ,  \label{eq:E_psi_K,nj}
\end{eqnarray}
where $\chi _{n}(\xi )$ is a harmonic-oscillator eigenfunction.

The connection between the small and large box regimes is given by
\begin{equation}
\lim\limits_{\lambda \rightarrow \infty }\lambda
^{-1/2}E_{m_{1},m_{2}m_{3},m_{4}}(\lambda )=2\left(
2n_{1}+2n_{2}+2n_{3}+3\right) .  \label{eq:lim_E}
\end{equation}
When carrying out this limit we should take into account that the symmetry
of the eigenfunction should be conserved as $\lambda \rightarrow \infty $
and that $\exp \left( iK\xi _{4}\right) $ does not exhibit a definite
symmetry. Therefore, in order to make a connection between both regimes the
eigenfunctions in the large-box case should be of the form
\begin{eqnarray}
\psi _{K,n_{1},n_{2},n_{3}}^{c}(\xi _{1},\xi _{2},\xi _{3},\xi _{4}) &=&\cos
\left( K\xi _{4}\right) \chi _{n_{1}}(\xi _{1})\chi _{n_{2}}(\xi _{2})\chi
_{n_{3}}(\xi _{3}),  \nonumber \\
\psi _{K,n_{1},n_{2},n_{3}}^{s}(\xi _{1},\xi _{2},\xi _{3},\xi _{4}) &=&\sin
\left( K\xi _{4}\right) \chi _{n_{1}}(\xi _{1})\chi _{n_{2}}(\xi _{2})\chi
_{n_{3}}(\xi _{3}).  \label{eq:psi_large_box}
\end{eqnarray}

In order to determine the symmetry of a given eigenfunction in the large-box
limit we take into account that $\xi _{4}$ is basis for the irrep $A_{1u}$
while the set $\{\xi _{1},\xi _{2},\xi _{3}\}$ is a basis for the irrep $%
T_{2u}$. Therefore, $\cos \left( K\xi _{4}\right) $ and $\sin \left( K\xi
_{4}\right) $ are basis for the irreps $A_{1g}$ and $A_{1u}$, respectively.
By means of the direct product of irreps we obtain the symmetry of any
eigenfunction of the form (\ref{eq:psi_large_box}). For example, when $%
\{n_{1},n_{2},n_{3}\}_{P}=\{0,0,1\}_{P}$ the three possible products $\chi
_{n_{1}}(\xi _{1})\chi _{n_{2}}(\xi _{2})\chi _{n_{3}}(\xi _{3})$ are basis
for $T_{2u}$ and the resulting functions $\psi
_{K,n_{1},n_{2},n_{3}}^{c}(\xi _{1},\xi _{2},\xi _{3},\xi _{4})$ and $\psi
_{K,n_{1},n_{2},n_{3}}^{s}(\xi _{1},\xi _{2},\xi _{3},\xi _{4})$ are basis
for $T_{2u}$ and $T_{2g}$, respectively.

\section{Simple variational method}

\label{sec:variational}

In order to obtain accurate variational results for large values of $\lambda
$ as was done in the case of two particles\cite{AF10} we should try
variational functions of the form
\begin{equation}
F(q_{1},q_{2},q_{3},q_{4})=G(q_{1},q_{2},q_{3},q_{4})\exp \left[ -a\left(
\xi _{1}^{2}+\xi _{2}^{2}+\xi _{3}^{2}\right) \right] ,
\label{eq:psi_var_good}
\end{equation}
where $G(q_{1},q_{2},q_{3},q_{4})$ satisfies the boundary conditions at the
box walls and $\exp \left[ -a\left( \xi _{1}^{2}+\xi _{2}^{2}+\xi
_{3}^{2}\right) \right] $ provides the correct asymptotic behaviour of the
wavefunction of the free oscillator. The variational parameter $a$ will
increase from $a_{0}$ to infinity as $\lambda $ increases from zero to
infinity.

However, since this calculation is rather cumbersome here we try a much
simpler one with a variational function of the form
\begin{equation}
F(q_{1},q_{2},q_{3},q_{4})=G(q_{1},q_{2},q_{3},q_{4})\exp \left[ -a\left(
q_{1}^{2}+q_{2}^{2}+q_{3}^{2}+q_{4}^{2}\right) \right] ,
\label{eq:psi_var_wrong}
\end{equation}
where we clearly sacrifice the correct description of the asymptotic
behaviour when $\lambda \rightarrow \infty $. For example,
\begin{eqnarray}
F_{A_{1g}}(q_{1},q_{2},q_{3},q_{4}) &=&N\left( q_{1}^{2}-1\right) \left(
q_{2}^{2}-1\right) \left( q_{3}^{2}-1\right) \left( q_{4}^{2}-1\right) \times
\nonumber \\
&&\times \exp \left[ -a\left( q_{1}^{2}+q_{2}^{2}+q_{3}^{2}+q_{4}^{2}\right)
\right] ,
\end{eqnarray}
\begin{eqnarray}
F_{A_{1u}}(q_{1},q_{2},q_{3},q_{4}) &=&N\xi _{4}\left( q_{1}^{2}-1\right)
\left( q_{2}^{2}-1\right) \left( q_{3}^{2}-1\right) \left(
q_{4}^{2}-1\right) \times  \nonumber \\
&&\times \exp \left[ -a\left( q_{1}^{2}+q_{2}^{2}+q_{3}^{2}+q_{4}^{2}\right)
\right] ,
\end{eqnarray}
and
\begin{eqnarray}
F_{T_{2u}}(q_{1},q_{2},q_{3},q_{4}) &=&N\left( q_{1}^{2}-1\right) \left(
q_{2}^{2}-1\right) \left( q_{3}^{2}-1\right) \left( q_{4}^{2}-1\right) \times
\nonumber \\
&&\times \exp \left[ -a\left( q_{1}^{2}+q_{2}^{2}+q_{3}^{2}+q_{4}^{2}\right)
\right] \left\{
\begin{array}{c}
\xi _{1} \\
\xi _{2} \\
\xi _{3}
\end{array}
\right.
\end{eqnarray}
are expected to yield approximations to the first energy levels $%
E_{1A_{1g}}<E_{1A_{1u}}<E_{1T_{2u}}$.

\begin{figure}[]
\begin{center}
\includegraphics[width=9cm]{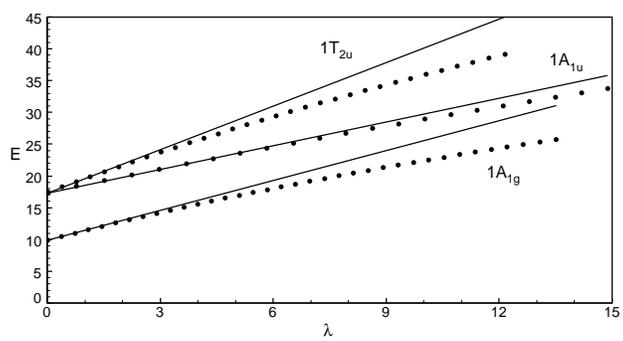}
\end{center}
\caption{Lowest three eigenvalues calculated by means of perturbation theory
(solid line) and the variational method (circles)}
\label{fig:EPTVAR}
\end{figure}

Figure~\ref{fig:EPTVAR} shows the perturbation and variational results for
those three states. Both the perturbation corrections of first order and the
variational approaches are upper bounds to the corresponding energies
because the variational principle applies to the lowest state of each
symmetry. Since the perturbation expressions shown in section~\ref
{sec:small-box} ($E_{PT}$ from now on) yield the exact result when $\lambda
=0$ they are expected to be more accurate than the variational results ($%
E_{var}$) for sufficiently small $\lambda $. Figure~\ref{fig:EPTVAR} reveals
that $E_{var}<E_{PT}$ for $\lambda >\lambda _{c}$ that tells us that $%
E_{var} $ is more accurate for sufficiently large values of $\lambda $. As
expected, the perturbation expressions exhibit a wrong behaviour for large
values of $\lambda $. On the other hand, $E_{var}$ increases correctly as $%
\sqrt{\lambda }$ but the choice of an incorrect exponential factor in the
trial function leads to a wrong coefficient of the leading term of the
energy; for example a numerical calculation suggests that
\begin{equation}
\lim\limits_{\lambda \rightarrow \infty }\lambda ^{-1/2}E_{1A_{1g}}=\sqrt{48}%
>6.
\end{equation}

\section{Conclusions}

\label{sec:conclusions}

Throughout this paper, as well as in the two earlier ones\cite{AF10,AF15},
we have shown that group theory is useful for the analysis of systems of
particles in a one-dimensional box. In the present case we can label the
states of the system of four particles by means of the irreps of the point
group $O_{h}$. The knowledge of the symmetry of the states for finite $%
\lambda $ and for $\lambda \rightarrow \infty $ facilitates the analysis of
the connection between the states of the confined and free systems,
respectively. In addition to it, point group proves suitable for the
construction of simple variational trial functions like those in section~\ref
{sec:variational} as well as for the construction of symmetry-adapted basis
sets for more accurate calculations like the Rayleigh-Ritz method used in
section~\ref{sec:small-box}.

\section*{Appendix: Construction of projection operators}

In this appendix we outline the procedure for the construction of the
projection operators that enabled us to determine the symmetry of the
corrections of first order to the eigenfunctions as well as to construct
symmetry-adapted basis sets and the variational trial functions.

First, we build a set of 48 matrices $G_{M}=\left\{ \mathbf{M}%
_{j},\;j=1,2,\ldots ,48\right\} $ given by the 24 permutations of the rows
of the $4\times 4$ identity matrix $\mathbf{I}$ and the 24 permutations of
the rows of $-\mathbf{I}$. This set of matrices is a group with respect to
the matrix product $\mathbf{M}_{i}\cdot \mathbf{M}_{j}$. Second, we define a
set $G_{O}=\left\{ O_{j},\;j=1,2,\ldots ,48\right\} $ of linear invertible
operators according to the rule
\begin{equation}
O_{j}f(\mathbf{q})=f(\mathbf{M}_{j}^{-1}\mathbf{q}),  \label{eq:O_j}
\end{equation}
where $\mathbf{q}$ is a column vector with elements $q_{i}$ (the four
dimensionless particle coordinates) and $f(\mathbf{q})$ is an arbitrary
function of them. These operators form a group with respect to the
composition $O_{i}\circ O_{j}$. If $\mathbf{M}_{i}$ and $\mathbf{M}_{j}$ are
the matrix representations of $O_{i}$ and $O_{j}$, respectively, then $%
\mathbf{M}_{i}\cdot \mathbf{M}_{j}$ is the matrix representation of $%
O_{i}\circ O_{j}$. In other words, the groups $G_{M}$ and $G_{O}$ are
isomorphic.

Third, we determine the classes for the group of matrices $G_{M}$ in the
usual way. Two matrices $\mathbf{M}_{i}$ and $\mathbf{M}_{j}$ belong to the
same class if $\mathbf{M}_{k}\cdot \mathbf{M}_{i}\cdot \mathbf{M}_{k}^{-1}=%
\mathbf{M}_{j}$ for some $\mathbf{M}_{k}\in G_{M}$. Fourth, we calculate the
traces, determinants and orders of the matrices in every class. In this way
we connect each class of matrices $\mathbf{M}_{j}$ (or operators $O_{j}$)
with the corresponding class of symmetry operations that appear in the
character table of the group $O_{h}$\cite{C90}.

Finally, we obtain the projection operators by means of the well known
expression\cite{H62,T64,C90}
\begin{equation}
P_{S}=\frac{n_{S}}{h}\sum_{j=1}^{h}\chi _{j}(S)O_{j},  \label{eq:P_S}
\end{equation}
where $h=48$ is the order of the group, $n_{S}$ is the dimension of the
irreducible representation $S$ and $\chi _{j}(S)$ is the character of the
operation $O_{j}$ for $S$ that appears in the character table. Since the
matrices in a class share the same trace it is sufficient to obtain a
one-to-one correspondence between the classes of matrices and the classes of
symmetry operations.

The application of the projection operators to the eigenfunctions of order
zero given by some of the sets of quantum numbers $\{n_{1},n_{2},n_{3},n_{4}%
\}_{P}$ yields the following irreps:

\[
\begin{array}{|l|c|c|l|}
\ \mathrm{Quantum\,numbers} & \mathrm{Number\,of\,states} & E_{0} & \mathrm{%
Symmetry} \\ \hline
\{1,1,1,1\} & 1 & \pi ^{2} & 1A_{1g} \\
\{1,1,1,2\}_{P} & 3 & 7\pi ^{2}/4 & 1A_{1u},\,1T_{2u} \\
\{1,1,2,2\}_{P} & 6 & 5\pi ^{2}/2 & 2A_{1g},\,1T_{2g},\,1E_{g} \\
\{1,1,1,3\}_{P} & 4 & 3\pi ^{2} & 3A_{1g},\,2T_{2g} \\
\{1,2,2,2\}_{P} & 4 & 13\pi ^{2}/4 & 2A_{1u},\,2T_{2u} \\
\{1,1,2,3\}_{P} & 12 & 15\pi ^{2}/4 & 3A_{1u},\,3T_{2u},\,1E_{u},\,4T_{2u},%
\,1T_{1u} \\
\{2,2,2,2\}_{P} & 1 & 4\pi ^{2} & 4A_{1g} \\
\{1,2,2,3\}_{P} & 12 & 9\pi ^{2}/2 & 5A_{1g},\,3T_{2g},\,4T_{2g},\,2E_{g},%
\,1T_{1g} \\
\{1,1,1,4\}_{P} & 4 & 19\pi ^{2}/4 & 4A_{1u},\,5T_{2u} \\
\{1,1,3,3\}_{P} & 6 & 5\pi ^{2} & 6A_{1g},\,3E_{g},\,5T_{2g} \\
\{2,2,2,3\}_{P} & 4 & 21\pi ^{2}/4 & 5A_{1u},\,6T_{2u} \\
\{1,1,2,4\}_{P} & 12 & 11\pi ^{2}/2 & 7A_{1g},\,2T_{1g},\,6T_{2g},\,4E_{g},%
\,2T_{1g}
\end{array}
\]

\end{document}